\documentclass[prl,preprint,superscriptaddress,showpacs]{revtex4}

\usepackage{amsmath}
\usepackage{amssymb}
\usepackage{graphicx}
\usepackage{subfigure}

\newcommand{\be}[1]{\begin{equation}\label{#1}}
\newcommand{\ee}{\end{equation}}
\newcommand{\ba}[1]{\begin{eqnarray}\label{#1}}
\newcommand{\ea}{\end{eqnarray}}
\newcommand{\rf}[1]{(\ref{#1})}
\newcommand{\nn}{\nonumber}

\newcommand{\diag}{\mbox{\rm diag}\,}

\begin{document}

%\title{$\mathcal{PT}$-symmetry and stability}
\title{$\mathcal{PT}$-symmetry, indefinite damping and dissipation-induced instabilities}

\author{Oleg N. Kirillov}
\affiliation{Helmholtz-Zentrum Dresden-Rossendorf,
P.O. Box 510119, D-01314 Dresden, Germany\\
Tel.: +49 351 260 2154, Fax: 2007, 12168,
E-mail: o.kirillov@hzdr.de}

\date{\today}

\begin{abstract}
%When gain and loss are in perfect balance, dynamical systems with indefinite damping can obey the exact $\mathcal{PT}$-symmetry and therefore be marginally stable with a pure imaginary spectrum.  At an exceptional point where the exact $\mathcal{PT}$-symmetry is spontaneously broken, the stability is lost via a Krein collision of eigenvalues just as it happens at the Hamiltonian Hopf bifurcation. In the parameter space of a general dissipative system, marginally stable $\mathcal{PT}$-symmetric ones occupy singularities on the boundary of the asymptotic stability domain. To observe how the singular surface governs dissipation-induced destabilization of the $\mathcal{PT}$-symmetric system when gain and loss are not matched, an extension of recent experiments with $\mathcal{PT}$-symmetric LRC circuits is proposed.
With perfectly balanced gain and loss, dynamical systems with indefinite damping can obey the exact $\mathcal{PT}$-symmetry being marginally stable with a pure imaginary spectrum.  At an exceptional point where the symmetry is spontaneously broken, the stability is lost via passing through a non-semisimple $1:1$ resonance. In the parameter space of a general dissipative system, marginally stable $\mathcal{PT}$-symmetric ones occupy singularities on the boundary of the asymptotic stability. To observe how the singular surface governs dissipation-induced destabilization of the $\mathcal{PT}$-symmetric system when gain and loss are not matched, an extension of recent experiments with $\mathcal{PT}$-symmetric LRC circuits is proposed.

\end{abstract}


\maketitle

\emph{Introduction.}
The notion of $\mathcal{PT}$-symmetry entered modern physics mainly from the side of quantum mechanics.
Parametric families of non-Hermitian Hamiltonians having both parity $(\mathcal{P})$ and time-reversal $(\mathcal{T})$ symmetry, possess pure real spectrum in some regions of the parameter space, which questions need for the Hermiticity axiom in quantum theory \cite{Bender1,Bender2,MB,Bender3}. First experimental evidence of $\mathcal{PT}$-symmetry and its violation came, however, from classical optics in media with inhomogeneous in space gain and damping \cite{prl2009,nature2010} and electrodynamics \cite{CircuitPT2011}.

$\mathcal{PT}$-symmetric equations of two coupled ideal LRC circuits, one with gain and another with loss,  have the form
\be{pt1}
\ddot {\bf z}+{\bf D}\dot {\bf z}+{\bf K}{\bf z}=0,
\ee
where dot stands for time differentiation and the real matrix of potential forces is ${\bf K}={\bf K}^T>0$
while the real matrix ${\bf D}={\bf D}^T$ of the damping forces is indefinite \cite{CircuitPT2011}.

For the problem considered in \cite{CircuitPT2011}, we assume that
\be{pt2}
{\bf D}={\bf D}_{\mathcal{PT}}=\left(
                                                                                       \begin{array}{rr}
                                                                                         -\delta  & 0  \\
                                                                                         0 & \delta  \\
                                                                                       \end{array}
                                                                                     \right),~~
{\bf K}={\bf K}_{\mathcal{PT}}=\left(
                                                                                       \begin{array}{rr}
                                                                                         k & \kappa  \\
                                                                                         \kappa & k \\
                                                                                       \end{array}
                                                                                     \right),
\ee
${\bf z}^T=(z_1,z_2)$, and $\delta$, $\kappa$ and $k$ are non-negative parameters. Eigenvalues of  ${\bf D}_{\mathcal{PT}}$ have equal absolute values and differ by sign, indicating perfect gain/loss balance in system \rf{pt1} with matrices \rf{pt2}.
The coordinate change $x_1=z_1+i z_2$, $x_2={x}_1^*$, $x_3=\dot x_1$, and $x_4=\dot x_2$,
where $i=\sqrt{-1}$ and the asterisk denotes complex conjugation, reduces this system to $i \dot {\bf x}={\bf H}{\bf x}$, where the Hamiltonian
\be{pt3}
{\bf H}=\left(
          \begin{array}{cccc}
            0 & 0 & i & 0 \\
            0 & 0 & 0 & i \\
            -ik & \kappa & 0 & i \delta \\
            -\kappa & -ik & i\delta & 0 \\
          \end{array}
        \right)
\ee
is $\mathcal{PT}$-symmetric (${\bf P}{\bf H}^*{=}{\bf H}{\bf P}$, ${\bf P}{=}\diag(1,-1,-1,1)$) \cite{W04, Bender4}.

In real electrical networks, additional losses may result in the indefinite damping matrices that possess both positive and negative eigenvalues with non-equal absolute values. A systematic study of dynamical systems \rf{pt1} with such a general \emph{indefinite damping}, has been initiated in \cite{Freitas1, Freitas2} in the context of distributed parameter control theory and  population biology \cite{Chen91,Freitas3,Joly}. In \cite{Kliem, Kirillov09,Damm11} gyroscopic stabilization of system \rf{pt1} was considered,
because negative damping produced by the falling dependence of the friction coefficient on the sliding velocity, feeds vibrations
in rotating elastic continua in frictional contact, e.g. in the singing wine glass \cite{Spurr, Akay, Kirillov08, Kirillov09a}.
In \cite{KPLA11} a gyroscopic $\mathcal{PT}$-symmetric system with indefinite damping was shown to originate in the studies of modulational instability of a traveling wave solution of the nonlinear Schr\"odinger equation (NLS) \cite{BD07}.
In nonlinear optics, a challenging problem of stability of localized solutions (solitons) is related to the indefinite damping, because stable
pulses in dual-core systems frequently exist far from the conditions that provide a perfect matching of gain and loss ($\mathcal{PT}$-symmetry) \cite{ Malomed96a,Malomed96b}. Recent techniques proposed for the stabilization of the solitons in two coupled perturbed NLSs include introduction of $\mathcal{PT}$-symmetric nonlinear gain and loss \cite{Kivshar2011} which signs can be periodically switched  \cite{Abdullaev2011,Malomed2011}.  Therefore, indefinite damping is a basic model to study how a localized supply of energy modifies the dissipative structure of a system \cite{Joly}.

In general, the eigenvalues $(\lambda)$ of system \rf{pt1}, when it is assumed that ${\bf z}\sim \exp(\lambda t)$, are complex with positive or negative real parts corresponding
either to growing or decaying in time solutions, respectively. \emph{Asymptotic stability} means decay of all modes.

A two-dimensional system \rf{pt1} with ${\bf D}=\delta \widetilde {\bf D}$ is asymptotically stable if and only if ${\rm tr}{\bf \widetilde D}>0$ and $0<\delta^2<\delta_{cr}^2$,
\be{pt4}
\delta_{cr}^2=\frac{({\rm tr}{\bf K \widetilde D}-\sigma_1({\bf K}){\rm tr}{\bf \widetilde D})({\rm tr}{\bf K \widetilde D}-\sigma_2({\bf K}){\rm tr}{\bf \widetilde D})}{-\det {\bf \widetilde D}{\rm tr} {\bf \widetilde D}({\rm tr}{\bf K \widetilde D}-{\rm tr}{\bf K}{\rm tr}{\bf \widetilde D}) },
\ee
where $\sigma_1({\bf K})$ and $\sigma_2({\bf K})$ are eigenvalues of ${\bf K}$ \cite{Freitas1, Kirillov2}.
However, when simultaneously ${\rm tr} {\bf \widetilde D}=0$ and ${\rm tr} {\bf K \widetilde D}=0$, the spectrum of the system \rf{pt1}
is \emph{Hamiltonian}, i.e. its eigenvalues are symmetric with respect to the imaginary axis of the complex plane \cite{Freitas2}. They are pure imaginary and simple (\emph{marginal stability}) if and only if $\delta^2 < \delta_{\mathcal{PT}}^2$,
\be{pt5}
\delta_{\mathcal{PT}}={\left|\sqrt{\sigma_1({\bf K})}-\sqrt{\sigma_2({\bf K})}\right|}\left({-\det{\bf \widetilde D}}\right)^{1/2}.
\ee

How the marginal stability domain of a indefinitely damped $\mathcal{PT}$-symmetric system relates to the domain
of asymptotic stability of a nearby dissipative system without this symmetry?
The answer is counterintuitive already for the thresholds \rf{pt4} and \rf{pt5}.
Our Letter describes mutual location of the two sets, thus linking the fundamental
concepts of modern physics: $\mathcal{PT}$-symmetry \cite{Bender1, Bender2, MB, Bender3} and dissipation-induced instabilities \cite{KMK,K04,KV10}.

\emph{A potential system with indefinite damping.}
First, we  extend the model \rf{pt1} with matrices \rf{pt2} by choosing the matrices of damping and potential forces in the form
\be{pt6}
{\bf D}=\left(
\begin{array}{ll}
\delta_1 & 0  \\
0 & \delta_2 \\
\end{array}
\right),~~
{\bf K}=\left(
\begin{array}{ll}
k_1 & \kappa  \\
\kappa & k_2 \\
\end{array}
\right),
\ee
where parameters can take arbitrary positive and negative values. For asymptotic stability it is necessary that
${\rm tr}{\bf D}>0$ and $\det{\bf K}>0$ \cite{Kirillov2}.

    \begin{figure}
    \begin{center}
    \includegraphics[width=0.8\textwidth]{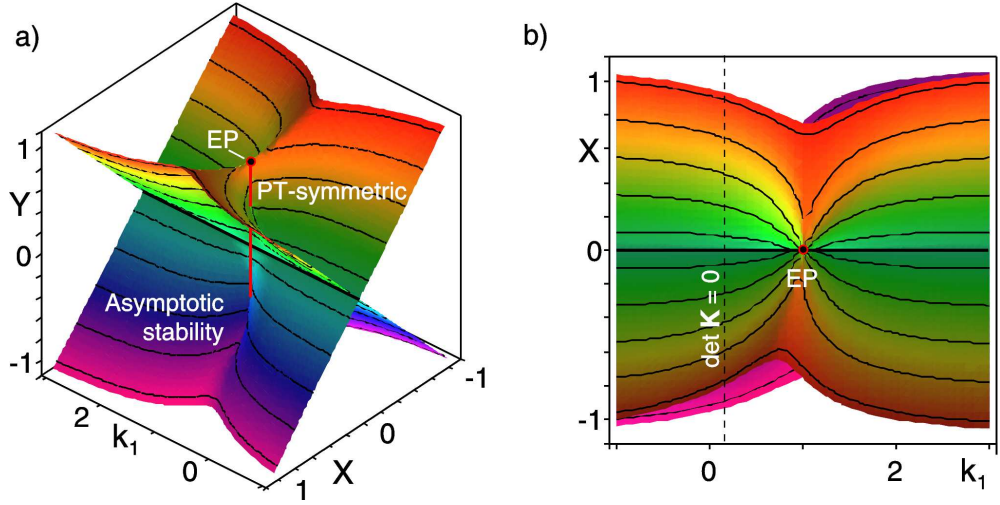}
    \end{center}
    \caption{(a) In the half-space $X>0$ of the $(k_1,X,Y)$ space, where $X=\delta_1+\delta_2$ and $Y=\delta_1-\delta_2$, a part of the singular surface locally equivalent to the Pl\"ucker conoid of degree $n=1$, bounds the domain of asymptotic stability of system \rf{pt1} with matrices \rf{pt6} and $\kappa=0.4$ and $k_2=1$; $\mathcal{PT}$-symmetric marginally stable systems occupy the red interval of self-intersection with two exceptional points (EPs) (black dots) at its ends. (b) The top view of the surface. }
    \label{fig2}
    \end{figure}

Introducing the parameters $X=\delta_1+\delta_2$ and $Y=\delta_1-\delta_2$,
we use the Routh-Hurwitz stability threshold \rf{pt4} where one should equate the right hand side to unity and replace the matrix
${\bf \widetilde D}$ with that given in Eq.~\rf{pt6}. The result is a quadratic equation for $k_1$. Expanding $k_1(X)$ in the vicinity of $X=0$, yields a linear approximation to the threshold of asymptotic stability
in the $(k_1,X)$ plane
\be{pt7}
k_1{=}k_2{+}\frac{1}{4}\frac{X}{Y}\left[Y^2\pm{\sqrt{\left(Y^2{-}{Y_{\mathcal{PT}}^-}^2\right)\left(Y^2{-}{Y_{\mathcal{PT}}^+}^2\right)}}\right].
\ee

$
Y_{\mathcal{PT}}^{\pm}=2\left(\sqrt{\sigma_2({\bf K})} \pm \sqrt{\sigma_1({\bf K})}\right),
$
where $\sigma_{1}=k_2-\kappa$ and $\sigma_{2}=k_2+\kappa$ are eigenvalues of the matrix $\bf K$ from Eq.~\rf{pt6} in which $k_1=k_2$ that happens when  $X=0$, i.e. $\delta_1=-\delta_2$. Therefore, on the line defined by the equations $k_1=k_2$ and $X=0$ in the $(k_1,X,Y)$ space, system \rf{pt1} with the matrices \rf{pt6} is reduced to the $\mathcal{PT}$-symmetric system with matrices \rf{pt2}
that is marginally stable on the interval $-Y_{\mathcal{PT}}^-<Y<Y_{\mathcal{PT}}^-$, cf. Eq.~\rf{pt5}.

        \begin{figure}
%        \labellist
%\pinlabel $X=0.2$ [r] at 210 206
%\pinlabel $k_1=1.05$ [r] at 595 246
%\pinlabel $X=0.05$ [r] at 595 206
%\endlabellist
    \begin{center}
    \includegraphics[width=0.8\textwidth]{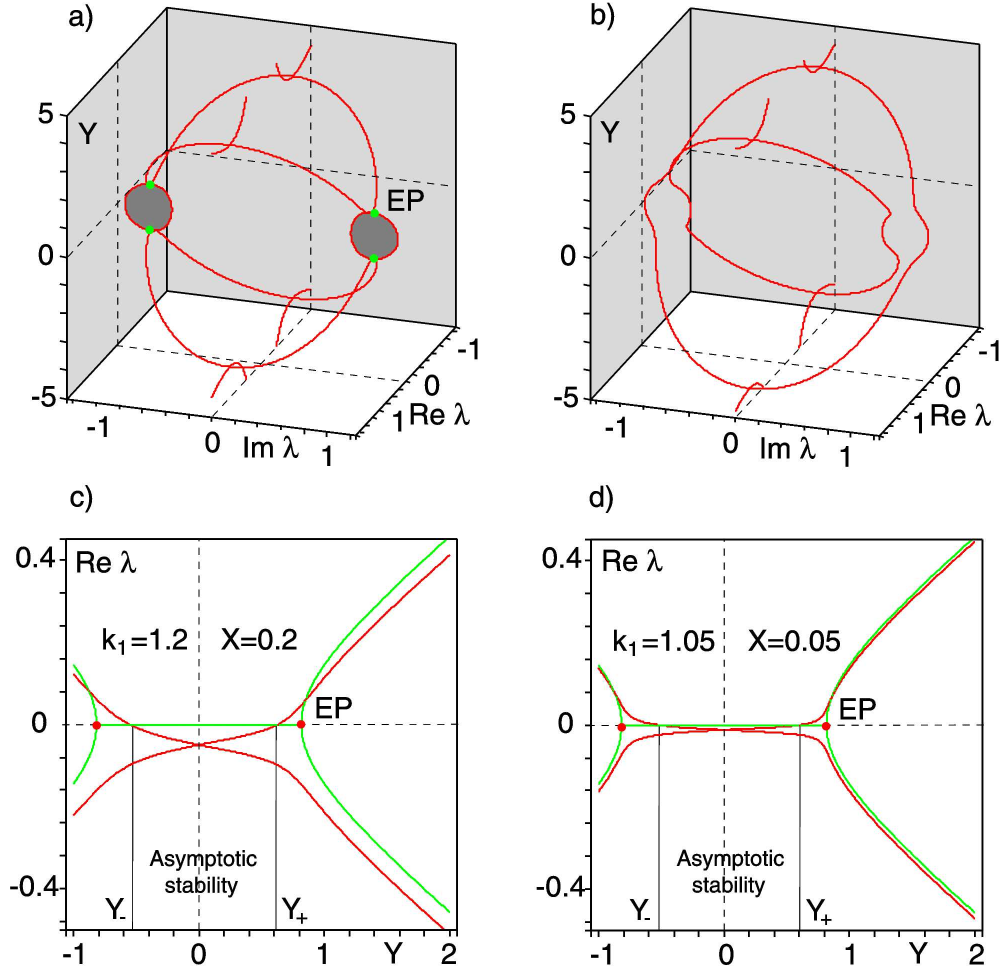}
    \end{center}
    \caption{Evolution of eigenvalues of system \rf{pt1} with matrices \rf{pt6} where $\kappa=0.4$ and $k_2=1$. (a) The loops of pure imaginary eigenvalues (dark grey) between the EPs marked by green dots imply marginal stability of the $\mathcal{PT}$-symmetric system corresponding to $k_1=1$ and $X=0$.
    (b) Unfolding the EPs of the unbalanced dissipative system with $k_1=1.2$ and $X=0.2$ and
    (c,d) its growth rates as functions of $Y$ (red curves). The growth rates vanish at the lower values of $Y$ not converging to the locations of the EPs of the $\mathcal{PT}$-symmetric system (red dots on a green curve) when $k_1\rightarrow 1$ and $X\rightarrow 0$ along a ray in the $(k_1,X)$ plane (the destabilization paradox \cite{Kirillov2,K04,KV10,Bottema}).   }
    \label{fig1}
    \end{figure}

In Fig.~\ref{fig2}(a) the vertical red line denotes this  interval with $Y_{\mathcal{PT}}^-\simeq0.817$ calculated for $k_2=1$ and $\kappa=0.4$. Along it $\mathcal{PT}$-symmetry is \emph{exact}, i.e. eigenvectors are also $\mathcal{PT}$-symmetric \cite{Bender1, Bender2, MB, Bender3}. Hence, the spectrum is pure imaginary, see Fig.~\ref{fig1}.
The ends of the interval are \emph{exceptional points} (EPs) \cite{Dietz} corresponding to the merging of a pair of pure imaginary eigenvalues into a double one with the Jordan block.
Passing through these points of the non-semisimple $1:1$ resonance with the increase of $|Y|$ is accompanied by the spontaneous breaking of the $\mathcal{PT}$-symmetry of eigenvectors although the system still obeys the symmetry. This causes  bifurcation of the double pure imaginary eigenvalues into complex ones with negative and positive real parts and oscillatory instability or \emph{flutter} when ${Y_{\mathcal{PT}}^-}^2<Y^2<{Y_{\mathcal{PT}}^+}^2$, see Fig.~\ref{fig1}(a). The bifurcation at $Y^2={Y_{\mathcal{PT}}^+}^2$ makes all the eigenvalues real of both signs (static instability or \emph{divergence}).

What happens with the stability near the red line in Fig.~\ref{fig2}(a)?
Fig.~\ref{fig1}(b) shows that, e.g. at the fixed $k_1=1.2$ and $X=0.2$, the eigencurves connected  at the EPs with  $Y=\pm Y_{\mathcal{PT}}^-$ in Fig.~\ref{fig1}(a), unfold into two non-intersecting loops in the $({\rm Re}\lambda, {\rm Im}\lambda, Y)$ space, manifesting an \emph{imperfect merging of modes} \cite{HG03} owing to gain/loss imbalance.

Now the stability is lost not via the passing through the non-semisimple $1:1$ resonance but because of migration of a pair of simple complex-conjugate eigenvalues from the left- to right-hand side of the complex plane at $|Y|<Y_{\mathcal{PT}}^-\simeq 0.817$. For example, tending the parameters to the point $(1,0)$ in $(k_1,X)$ plane along a ray, specified by the equation $X=k_1-1$, we find that the thresholds of asymptotic stability converge to the limiting values of $Y_{+}\simeq 0.615 < 0.817$ and $Y_{-} \simeq -0.531 > -0.817$, see Fig.~\ref{fig1}(c,d). The limits vary with the change of the slope of the ray. Therefore, infinitesimal imperfections in the loss/gain balance and in the potential, destroying the $\mathcal{PT}$-symmetry, can significantly decrease the interval of asymptotic stability with respect to the marginal stability interval.

Such a paradoxical finite jump in the instability threshold caused by a tiny variation in the damping distribution, typically occurs in dissipatively perturbed autonomous Hamiltonian or reversible systems \cite{Kirillov2,MK91} of structural and contact mechanics \cite{HG03, KMK, KV10} and  hydrodynamics \cite{Romea77, KM09, Swaters10}, as well as in periodic non-autonomous ones \cite{HR95}. We have just described a similar effect when the marginally stable system is dissipative but obeys $\mathcal{PT}$-symmetry.

A reason for the dependence of the limiting critical value of $Y$ on the direction of approach follows from the linear approximation \rf{pt7}, which defines two straight lines orthogonal to the $Y$-axis. When  $Y$ changes from $-Y_{\mathcal{PT}}^-$ to $ Y_{\mathcal{PT}}^-$, the straight lines \rf{pt7} rotate around the $Y$-axis. We remind that a set of points swept by a moving straight line is called a \emph{ruled surface} \cite{HK10,BG88}. A \emph{right conoid} is a ruled surface generated by a family of straight lines that all intersect orthogonally a fixed straight line (the $Y$-axis in our case). Therefore, Eq.~\rf{pt7} defines a right conoid in the $(k_1,X,Y)$-space. In order to identify its type, we observe that Eq.~\rf{pt7} results in a cubic equation for $Y$. The third-degree term in it can be neglected when $|Y|<Y_{\mathcal{PT}}^-$. Resolving the remaining quadratic equation and introducing the polar coordinates $(\rho, \phi)$ in the $(k_1,X)$ plane as $k_1=k_2+\rho\cos\phi$ and $X=\frac{\rho\sin\phi}{\sqrt{k_2}}$, we find a parametric surface
\be{pt8}
(\rho,\phi)\mapsto \left(k_2+\rho\cos\phi,\frac{\rho\sin\phi}{\sqrt{k_2}},´\frac{2\kappa}{\sqrt{k_2}} \sin\phi \right).
\ee
This is a canonical equation for the special type of the right conoid known as the \emph{Pl\"ucker conoid} of degree 1 --- a singular surface with one horizontal and one vertical interval of self-intersection \cite{HK10,BG88}. The latter has at its ends two \emph{Whitney umbrella} singularities \cite{Bottema, Gils, Langford}.

    \begin{figure}
    \begin{center}
    \includegraphics[width=0.8\textwidth]{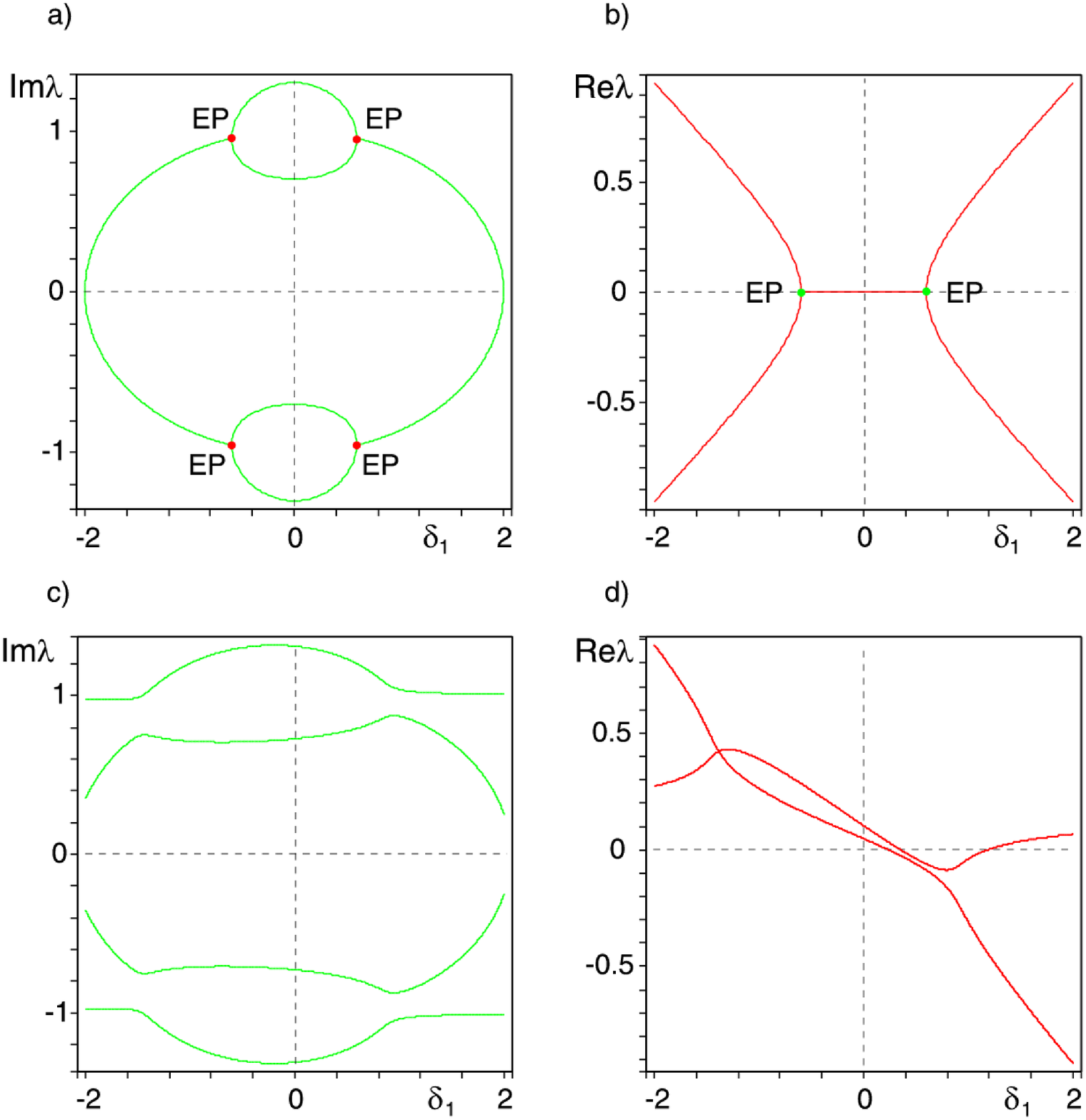}
    \end{center}
    \caption{ Imaginary and real parts of the eigenvalues of the gyroscopic system \rf{pt9} as functions of the damping parameter $\delta_1$  for $k_1=1$, $\Omega=0.3$ and (a,b)  $\kappa=0$, $\delta_2=-\delta_1$ ($\mathcal{PT}$-symmetric case), (c,d)  $\kappa=0.1$, $\delta_2=-0.3$.}
    \label{fig3}
    \end{figure}

Near the interval $-Y_{\mathcal{PT}}^-\le Y \le Y_{\mathcal{PT}}^-$ shown in red in Fig.~\ref{fig2}(a), the boundary of asymptotic stability given by Eq.~\rf{pt4} converges to the Pl\"ucker conoid \rf{pt7}, which is its exact linear approximation.
The latter, in turn, is approximated by the ruled surface \rf{pt8} that is in a canonical form for the Pl\"ucker conoid.
Qualitatively, all the three surfaces have the same singularities visible in Fig.~\ref{fig2}.

The approximation of type \rf{pt8} can also be obtained from the perturbation formulas for splitting double semi-simple eigenvalues $\pm i k_2$ (diabolical points) corresponding to $\kappa=0$, $k_1=k_2$ and $\delta_{1,2}=0$, see \cite{Kirillov08, Kirillov09a}. The Pl\"ucker conoid of degree 1 singularity on the boundary of the asymptotic stability domain generically occurs as a result of the unfolding of the semi-simple $1:1$-resonance \cite{HK10,Kirillov2}.

The $\mathcal{PT}$-symmetric marginally stable system studied in \cite{CircuitPT2011}, occupies a common `handle' of the two Whitney umbrellas on the Pl\"ucker conoid surface. The surface forms an instability threshold for  the nearby systems with the gain/loss mismatch and additional coupling in the matrix of potential forces. These imperfections are realizable in the physical LRC-circuits. This opens a way for the experimental investigation of dissipation-induced instabilities and related paradoxes that are common for very different dynamical systems \cite{Bottema, KMK}. Indeed,  since the singular geometry behind the destabilization paradox in dissipatively perturbed Hamiltonian, reversible, and $\mathcal{PT}$-symmetric systems is the same, the experiments with the near-$\mathcal{PT}$-symmetric LRC-circuits promise to be an efficient alternative to the mechanical ones. Development of the latter is restrained in particular by insufficient so far accuracy in damping identification.

    \begin{figure}
    \begin{center}
    \includegraphics[width=0.8\textwidth]{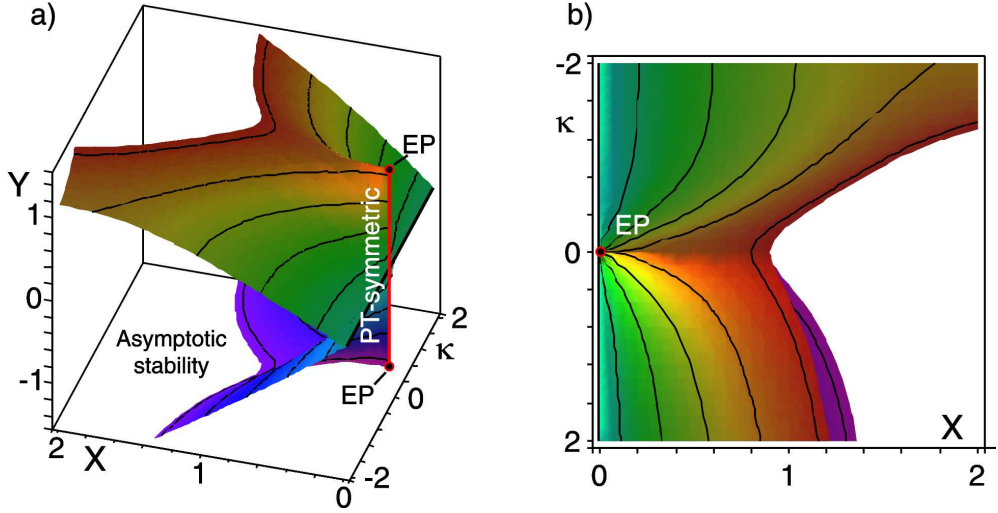}
    \end{center}
    \caption{(a) The domain of asymptotic stability and its boundary for the gyroscopic system \rf{pt9} in the $(\kappa,X,Y)$-space when $k_1=1$ and $\Omega=0.3$. The vertical red interval of self-intersection corresponds to the domain of marginal stability of the $\mathcal{PT}$-symmetric gyroscopic system with the indefinite damping.  (b) The top view of the stability boundary.}
    \label{fig4}
    \end{figure}

\emph{A gyroscopic system with indefinite damping.}
Taking into account commercial availability of \emph{gyrators}  --- the non-reciprocal elements of LRC circuits that model gyroscopic effects  \cite{Tellegen48,Fabre,Figotin,Potton} --- it should be possible to extend the experiments described in \cite{CircuitPT2011} to the gyroscopic systems with the indefinite damping \cite{Kliem}.

Consider a system with two degrees of freedom
\be{pt9}
\ddot {\bf z}+({\bf D}+2\Omega{\bf J})\dot {\bf z}+({\bf K}+(\Omega{\bf J})^2){\bf z}=0,
\ee
where ${\bf J}
$ is a matrix of gyroscopic forces with the entries $j_{11}=j_{22}=0$ and $j_{21}=-j_{12}=1$, $\Omega$ is a gyroscopic parameter, and
$\bf D$ and $\bf K$ are matrices of damping and potential forces.
Eq.~\rf{pt9} describes stability of a particle in a rotating saddle trap and flexible shafts in the classical rotor dynamics and arises in the theories of helical quadrupole magnetic focussing systems of accelerator physics and light propagation in liquid crystals \cite{Brouwer18,Bottema76,Chernin1983,Shapiro2001, KPLA11}.

When ${\bf D}=\diag(\delta_1, -\delta_1)$ and ${\bf K}=\diag(k_1,k_1)$, the system \rf{pt9} is invariant under transformations $t \leftrightarrow -t$ and $z_1\leftrightarrow z_2$, i.e. it is $\mathcal{PT}$-symmetric \cite{CircuitPT2011}.

Assume ${\bf D}=\diag(\delta_1,\delta_2)$ and ${\bf K}=\diag(k_1,k_1+\kappa)$.
In Fig.~\ref{fig3} we plot the imaginary and real parts of the eigenvalues as functions of $\delta_1$. When $\delta_2=-\delta_1$ and $\kappa=0$, the spectrum is symmetric with respect to the imaginary axis of the complex plane and demonstrates a typical for the  $\mathcal{PT}$-symmetric system behavior, see Fig.~\ref{fig3}(a,b). Detuning the gain and loss as well as the potential, unfolds the EPs and creates an interval of the asymptotic stability that is smaller than the interval of the marginal stability, see Fig.~\ref{fig3}(c,d).

With the parameters $X=\delta_1+\delta_2$, and $Y=\delta_1-\delta_2$, we plot the Routh-Hurwitz threshold for the asymptotic stability of system \rf{pt9} in the $(\kappa,X,Y)$ space in Fig.~\ref{fig4}. Again, the surface is locally equivalent to the Pl\"ucker conoid. $\mathcal{PT}$-symmetric marginally stable systems live on the vertical interval of self-intersection terminated by two exceptional points. The Whitney umbrella singularities at the EPs are responsible for the dissipation-induced enhancement of instability found in \cite{BD07}.

\emph{Example. Dissipatively enhanced modulational instability}

A monochromatic plane wave with a finite amplitude propagating in a dispersive
medium can be disrupted into a train of short pulses when the amplitude exceeds some threshold. This process develops due to an unbounded increase in the percentage modulation of the wave, i.e. instability of the carrier wave with respect to modulations. This is a fundamental for modern fluid dynamics, nonlinear optics and plasma physics \emph{modulational instability} \cite{ZO2009}. This instability, discovered by Bespalov and Talanov and Benjamin and Feir \cite{BespalovTalanov66,BF67}, can trigger formation of the breather-type solitons from the Stokes waves in deep water. The breathers are associated with the rogue waves, recently detected in a water wave tank  \cite{Hoffmann11prl}.

The modulational instability can be enhanced with additional dissipation \cite{BD07}. Below we show that this effect is rooted in the mutual location of $\mathcal{PT}$-symmetric gyroscopic systems with indefinite damping with respect to general dissipative ones.

Without dissipation, a slowly varying in time envelope $A$ of the rapidly oscillating carrier wave is often described
by the nonlinear Schr\"odinger equation (NLS)
\be{bf00}
iA_t+\alpha A_{xx}+\gamma|A|^2A=0,
\ee
where  $\alpha$ and $\gamma$ are
positive real numbers, $i=\sqrt{-1}$, and the modulations
are restricted to one space dimension $x$ \cite{BD07,ZO2009}.
Eq.~\rf{bf00} has a solution in the form of a monochromatic wave
\be{bf01}
A=A_0e^{ikx-i\omega t},
\ee
where the frequency of the modulation, $\omega$, depends on the amplitude $A_0=u_1^0+iu_2^0$ and spacial wavenumber $k$ as $\omega=\alpha k^2-\gamma \| {\bf u}_0 \|^2$ with ${\bf u}_0^T=(u_1^0,u_2^0)$.

We linearize the NLS about the basic traveling wave solution \rf{bf01} in order to study stability of the modulation. Assuming periodic in $x$ perturbations with the wavenumber $\sigma$ we substitute their Fourier expansions into the linearized problem. Then, the $\sigma$-dependent modes decouple into four-dimensional
subspaces for each harmonic with the number $n$, so that for $n=1$ we get \cite{BD07}
\ba{bf02}
{\bf J}\dot {\bf v}+2\alpha k \sigma {\bf J}{\bf w}-\alpha \sigma^2 {\bf v}+2\gamma {\bf u}_0{\bf u}_0^T{\bf v}&=&0,\nn\\
{\bf J}\dot {\bf w}-2\alpha k \sigma {\bf J}{\bf v}-\alpha \sigma^2 {\bf w}+2\gamma {\bf u}_0{\bf u}_0^T{\bf w}&=&0,
\ea
where dot indicates time differentiation and the dyad ${\bf u}_0{\bf u}_0^T$ is a $2 \times 2$ symmetric matrix. Eq.~\rf{bf02} can be transformed to that of the indefinitely damped gyroscopic system \rf{pt9} with $\Omega=\alpha \sigma^2 -\gamma \|{\bf u}_0 \|^2$, ${\bf D}=2\gamma({\bf u}_0{\bf u}_0^T{\bf J}-{\bf J}{\bf u}_0{\bf u}_0^T)$, and ${\bf K}=(4\alpha^2k^2\sigma^2+\gamma^2\|{\bf u}_0 \|^4){\bf I}$, where $\bf I$ is a unit matrix, which is $\mathcal{PT}$-symmetric because the eigenvalues $\pm 2\gamma \| {\bf u}_0\|^2$ of the matrix ${\bf D}$ differ by sign only \cite{KPLA11}. This implies that the spectrum of the system \rf{bf02} is Hamiltonian, i.e. symmetric with respect to both real and imaginary axis of the complex plane \cite{Freitas2,ZO2009,BD07}, with the eigenvalues
\be{bf03}
\lambda=\pm i 2\alpha k\sigma \pm i \sigma\sqrt{2\alpha \gamma(\| {\bf u}_0 \|_i^2- \| {\bf u}_0 \|^2)},
\ee
where
\be{bf04}
\| {\bf u}_0 \|_i^2=\frac{\alpha\sigma^2}{2\gamma}.
\ee
At small amplitudes of the modulation, the eigenvalues are pure imaginary. With the increase in the amplitude, the modes with the opposite Krein signature collide at the threshold $\| {\bf u}_0 \|=\| {\bf u}_0 \|_i$ \cite{BD07}.
At $\| {\bf u}_0 \|>\| {\bf u}_0 \|_i$ the double pure imaginary eigenvalue splits into complex-conjugate eigenvalues, one of which with positive real part, that corresponds to the modulational instability in the ideal (undamped) case \cite{ZO2009,BD07}.

    \begin{figure}[ftp]
    \begin{center}
    \includegraphics[width=0.97\textwidth]{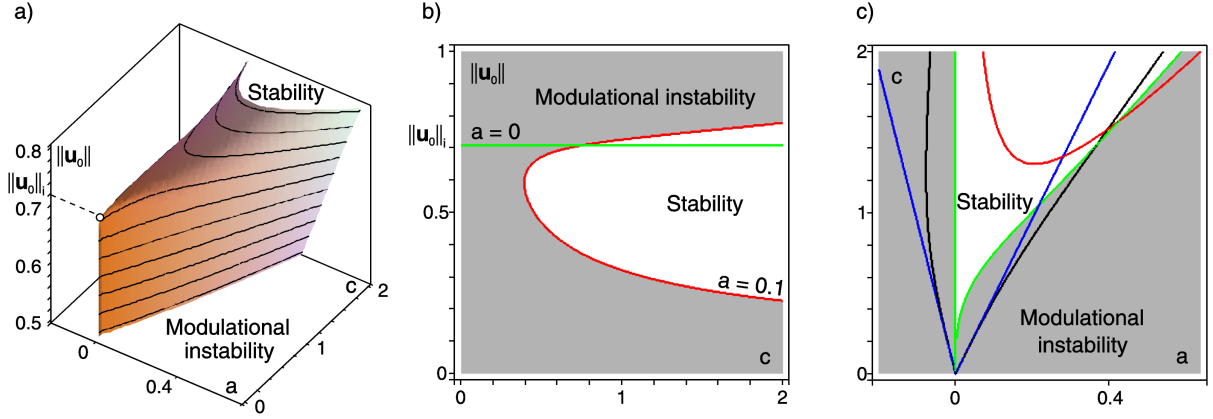}
    \end{center}
    \caption{(a) The boundary between the domains of asymptotic stability and  modulational instability in the $(a,c,\|{\bf u}_0 \|)$ space when $\sigma=1$, $\alpha=1$, $\gamma=1$, and $k=1$. (b) In $(c,\|{\bf u}_0 \|)$ plane the cross sections of the boundary at (green) $a=0$ and (red) $a=0.1$. (c) In $(a,c)$ plane the cross sections of the boundary at (green) $\|{\bf u}_0 \|=\|{\bf u}_0 \|_i=\frac{\sqrt{2}}{2}$, (black) $\|{\bf u}_0 \|=\|{\bf u}_0 \|_i-0.05$, and (red) $\|{\bf u}_0 \|=\|{\bf u}_0 \|_i+0.05$; blue lines is a linear approximation \rf{bf08}.}
    \label{fig5}
    \end{figure}

Introducing into Eq.~\rf{bf00} the dispersive and nonlinear losses with the coefficients $a$ and $c$, respectively, we arrive at the dissipatively-perturbed NLS \cite{Malomed96a,Malomed96b,BD07}
\be{bf05}
iA_t+(\alpha -i a)A_{xx}+(\gamma + i c)|A|^2A=0,
\ee
which after linearization and use of Fourier expansions yields the reduced system \cite{BD07}
\ba{bf06}
{\bf J}\dot {\bf v}+2\alpha k \sigma {\bf J}{\bf w}-\alpha \sigma^2 {\bf v}+2\gamma {\bf u}_0{\bf u}_0^T{\bf v}+2ka\sigma{\bf w}+a\sigma^2{\bf J}{\bf v}+2c{\bf u}_0^T{\bf v}{\bf J}{\bf u}_0&=&0,\nn\\
{\bf J}\dot {\bf w}-2\alpha k \sigma {\bf J}{\bf v}-\alpha \sigma^2 {\bf w}+2\gamma {\bf u}_0{\bf u}_0^T{\bf w}-2ka\sigma {\bf v}+a\sigma^2{\bf J}{\bf w}+2c{\bf u}_0^T{\bf w}{\bf J}{\bf u}_0&=&0.
\ea
When $a=0$ and $c=0$, Eqs.~\rf{bf06} coincide with the ideal system \rf{bf02}.

Writing the Routh-Hurwitz conditions for the characteristic polynomial of the system \rf{bf06}, we find an expression for the threshold of the modulational instability in the presence of dissipation
\ba{bf07}
2c^2(c a-\gamma\alpha)\| {\bf u}_0 \|^6+(4\sigma^2ca(ca-\gamma\alpha)-4a^2k^2(\gamma^2+c^2)+c^2\sigma^2(a^2+\alpha^2))\| {\bf u}_0 \|^4
&+&\\
2a\sigma^2(\alpha\sigma^2(\alpha c-\gamma a)+2\sigma^2ca^2+4ak^2(\gamma\alpha-ca))\| {\bf u}_0 \|^2+a^2\sigma^4(\sigma^2-4k^2)(a^2+\alpha^2)&=&0.\nn
\ea
The threshold equation \rf{bf07} yields a linear approximation to the stability boundary
in the $(a,c)$ plane of the coefficients of dispersive and nonlinear losses \cite{BD07}
\be{bf08}
c=\frac{\sigma}{\| {\bf u}_0 \|^2}\left[-{\sigma}\pm\frac{k(2\| {\bf u}_0 \|_i^2- \| {\bf u}_0 \|^2)}{\| {\bf u}_0 \|_i\sqrt{\| {\bf u}_0 \|_i^2- \| {\bf u}_0 \|^2}}\right] a+o(a).
\ee
When $a \ll c$, a simple approximation follows from Eq.~\rf{bf08} to the amplitude at the threshold of the modulational instability in the presence of dissipation
\be{bf09}
\| {\bf u}_0 \|_d\simeq\| {\bf u}_0 \|_i-\frac{1}{2}\frac{k^2\sigma^2}{\| {\bf u}_0 \|_i^3}\frac{a^2}{c^2}\le\| {\bf u}_0 \|_i.
\ee
Note that Eq.~\rf{bf09} is in the canonical for the Whitney umbrella form $Z=X^2/Y^2$ \cite{BG88}.

In Fig.~\ref{fig5}(a) the threshold \rf{bf07} is shown in the $(a,c,\| {\bf u}_0 \|)$ space. At $\| {\bf u}_0 \|=\| {\bf u}_0 \|_i$ and $a=0$ and $c=0$ it has the Whitney umbrella singularity at the exceptional point; along the interval $\| {\bf u}_0 \| \le \| {\bf u}_0 \|_i$ the system is $\mathcal{PT}$-symmetric with pure imaginary spectrum. Below the surface \rf{bf07} when $a>0$ and $c>0$ the dissipative system \rf{bf06} with the broken $\mathcal{PT}$-symmetry is asymptotically stable. In Fig.~\ref{fig5}(b) the cross-sections of the stability boundary \rf{bf07} are shown for $a=0$ (green line) and $a=0.1$ (red line) in the $(c,\| {\bf u}_0 \|)$ plane. The domain of modulational instability that was above the green line in Fig.~\ref{fig5}(b) when $a=0$ expands considerably below this line (grey area) when the coefficient of dispersive losses $a\ne 0$ (\emph{enhancement of the modulational instability with dissipation} \cite{BD07}). Fig.~\ref{fig5}(c) shows the cross-sections of the surface \rf{bf07} in the $(a,c)$ plane for $\| {\bf u}_0 \|=\| {\bf u}_0 \|_i$ (green line) and when $\| {\bf u}_0 \|$ is slightly above (red line) or below (black line) the amplitude at the threshold of the modulational instability in the undamped case. The cross-sections are typical for the surface with the Whitney umbrella singularity \cite{KV10}. In particular, they justify the approximation \rf{bf08} (blue lines) to the stability boundary that yields the canonical equation for the Whitney umbrella \rf{bf09}.

\emph{Summary.} A direct link is established between the $\mathcal{PT}$-symmetry and dissipation-induced instabilities: The systems with the exact $\mathcal{PT}$-symmetry are identified with the singularities on the threshold of asymptotic stability of the indefinitely damped ones.
This finding opens a new perspective for $\mathcal{PT}$-symmetric LRC circuit experiments that could test the both fundamental physical concepts, which is so far unavailable in the mechanical experiments. As an example, the enhancement of the modulational instability with dissipation is connected to the existence of the Whitney umbrella singularity on the instability threshold.

\end{document}